\documentclass[aps,prl,twocolumn,superscriptaddress,showpacs]{revtex4}
\usepackage{color}
\usepackage{graphicx}
\usepackage{dcolumn}
\usepackage{bm}

\begin{document}

\preprint{APS/123-QED}

\title{Dynamical Behavior of Spins in the Rare-Earth Kagom\'{e} Pr$_3$Ga$_5$SiO$_{14}$}

\author{L. L. ~Lumata}
\affiliation{Department of Physics  and National High Magnetic
Field Laboratory, Florida State University, Tallahassee, Florida
32310, USA}

\author{K.-Y. Choi}
\affiliation{Department of Chemistry  and National High Magnetic
Field Laboratory, Florida State University, Tallahassee, Florida
32310, USA}

\affiliation{Department of Physics, Chung-Ang University, 221
Huksuk-Dong, Seoul 156-756, Republic of Korea}

\author{T. Besara}
\affiliation{Department of Physics  and National High Magnetic
Field Laboratory, Florida State University, Tallahassee, Florida
32310, USA}

\author{M. J. R. Hoch}
\affiliation{Department of Physics and National High Magnetic
Field Laboratory, Florida State University, Tallahassee, Florida
32310, USA}

\author{H. D. Zhou}
\affiliation{Department of Physics  and National High Magnetic
Field Laboratory, Florida State University, Tallahassee, Florida
32310, USA}

\author{J. S. ~Brooks}
\affiliation{Department of Physics and National High Magnetic
Field Laboratory, Florida State University, Tallahassee, Florida
32310, USA}

\author{P. L. ~Kuhns}
\affiliation{Department of Physics  and National High Magnetic
Field Laboratory, Florida State University, Tallahassee, Florida
32310, USA}

\author{A. P. ~Reyes}
\affiliation{Department of Physics  and National High Magnetic
Field Laboratory, Florida State University, Tallahassee, Florida
32310, USA}

\author{N. S. Dalal}
\affiliation{Department of Chemistry  and National High Magnetic
Field Laboratory, Florida State University, Tallahassee, Florida
32310, USA}

\author{C. R. Wiebe}
\affiliation{Department of Physics  and National High Magnetic
Field Laboratory, Florida State University, Tallahassee, Florida
32310, USA}
\date{\today}

\begin{abstract}
We report on the use of $^{69,71}$Ga nuclear magnetic resonance to
probe spin dynamics in the rare-earth kagom\'{e} system
Pr$_3$Ga$_5$SiO$_{14}$. We find that the spin-lattice relaxation
rate $^{69}1/T_1$ exhibits a maximum around 30 K, below which the
Pr$^{3+}$ spin correlation time $\tau$ shows novel field-dependent
behavior consistent with a field-dependent gap in the excitation
spectrum. The spin-spin relaxation rate $^{69}1/T_{2}$ exhibits a
peak at a lower temperature (10 K) below which field-dependent
power-law behavior close to $T^{2}$ is observed. These results
point to field-induced formation of nanoscale magnetic clusters
consistent with recent neutron scattering measurements.
\end{abstract}


\maketitle


Various two-dimensional (2D) triangular lattices, including
kagom\'{e} systems, exhibit intriguing low-temperature spin
dynamics induced by a geometrical frustration~\cite{Diep,
Ramirez1, Helton, Ramirez, Wills, 3,4,5,6,7,8,9,10,11,12,13}. The
prototypical kagom\'{e} examples are ZnCu$_3$(OH)$_6$Cl$_{12}$
[Cu$^{2+}$ (S=1/2)], SrCr$_{8-x}$Ga$_{4+x}$O$_{19}$ [Cr$^{3+}$
(S=3/2)], and the jarosite (D$_3$O)Fe$_3$(SO$_4$)$_2$(OD)$_{6}$
[Fe$^{3+}$ (S=5/2)]~\cite{Helton,Ramirez,Wills}. These are all 3d
transition metal based compounds in which the exchange
interactions and geometrical frustration are of dominant
importance while Dzyaloshinsky-Moriya interactions, single-ion
anisotropies, and off-stoichiometry serve as small perturbations.
Nonetheless, investigation of the predicted spin-liquid state is
complicated by the presence of these perturbations.

For rare-earth (RE) compounds, the situation might be less
complicated: the single-ion anisotropies govern the overall
magnetic behavior, while frustration-induced spin dynamics is
expected to emerge at energies well below the crystal field
splitting. The recently discovered RE based kagom\'{e} compounds
R$_3$Ga$_5$SiO$_{14}$ (R=Nd or Pr) do show a disordered state at
low temperatures~\cite{14,15,16,17}. The cooperative magnetic
correlations depend on the RE ion involved and the crystal field
environment. However, this issue has not yet been fully addressed.
It is therefore desirable to study the spin dynamics of
Pr$_3$Ga$_5$SiO$_{14}$ and to compare the behavior with its
isostructural counterpart Nd$_3$Ga$_5$SiO$_{14}$ which has
different magneto-crystalline anisotropy \cite{17}.

\begin{figure}[tbp]
\linespread{1}
\par
\includegraphics[width=3.3in]{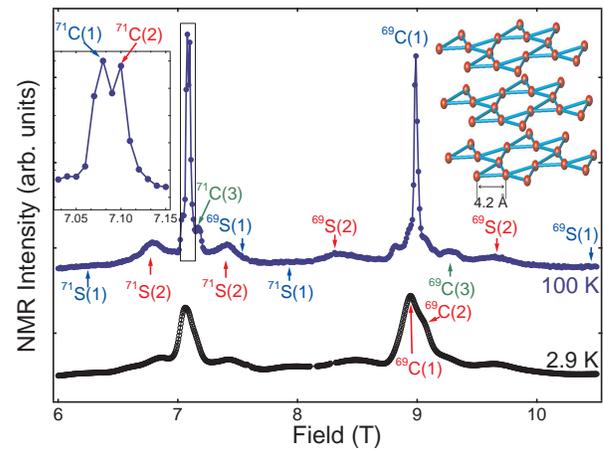}
\par
\caption{(color online). NMR field-scan spectra of
Pr$_3$Ga$_5$SiO$_{14}$ at 92 MHz showing quadrupolar split
$^{69}$Ga ($I = 3/2$) and $^{71}$Ga ($I = 3/2$) components for the
three non-equivalent Ga sites. Resolved peaks are labelled central
(C) or satellite (S) components. The left inset shows the enlarged
version of the portion of the graph in box. The right inset
depicts the crystal structure showing three kagom\'{e} planes. Two
Ga sites lie between these planes and the third site is in plane
(see Ref. \cite{17}).} \label{fig:1}
\end{figure}

Pr$_3$Ga$_5$SiO$_{14}$, which belongs to the langasite family, has
a trigonal crystal structure ($P321$) with lattice parameters
$\mathrm{a=8.0661(2) \AA}$ and $\mathrm{c=5.0620(2) \AA}$. The RE
Pr$^{3+}$ ions ($5f^2:J=4$) are networked by corner sharing
triangles in well-separated planes to form a distorted lattice
(see the inset of Fig.~\ref{fig:1}) topologically equivalent to
the kagom\'{e} lattice. Analysis of the specific heat suggests
that the crystal field energy level structure of Pr$^{3+}$
consists of three singlets with gaps $\Delta_1=25$~K and
$\Delta_2=68$ K, and $\Delta_3=780$~K~\cite{15}. At high
temperatures the magnetic susceptibility $\chi$ obeys the
Curie-Weiss law with a Curie-Weiss temperature of
$\mathrm{\theta_{CW}=-2.3}$~K that points to weak
antiferromagnetic (AF) correlations between the Pr$^{3+}$
spins~\cite{15}. At zero external field Pr$_3$Ga$_5$SiO$_{14}$
shows (i) no long-range magnetic ordering at temperatures down to
35~mK, (ii) a $T^2$ dependence of the specific heat at low
\emph{T}, and (iii) spin excitations consistent with a highly
degenerate state~\cite{15}. These suggest that
Pr$_3$Ga$_5$SiO$_{14}$ has a spin-liquid-like ground state
corresponding to a high frustration index
$\mathrm{f=\theta_{CW}/T_{c}\geq 66}$. Application of an external
field parallel to the \emph{c}-axis leads to the formation of
nanoscale magnetic clusters, whose size increases with \emph{H}.
This behavior is accompanied by reduction of the $T^{2}$ component
of the specific heat and the opening of a spin gap in the
excitation spectrum \cite{15}.

$\mu$SR and NMR measurements on the sister RE kagom\'{e} compound
Nd$_3$Ga$_5$SiO$_{14}$ reveal that a disordered state persists
down to low \emph{T} in zero field \cite{16,17}. Spin fluctuations
are suppressed by an applied field and for \emph{H} = 0.5 T a
field-induced transition is found at 60 mK. Although the two
isostructural compounds R$_3$Ga$_5$SiO$_{14}$ (R = Pr$^{3+}$ or
Nd$^{3+}$) share, to some extent, common physics, detailed
cooperative spin dynamics varies with the RE ions. This is because
Pr$^{3+}$ and Nd$^{3+}$ have different crystal field splittings,
single-ion anisotropies, and exchange interactions leading to a
substantial difference in low-energy spin dynamics both at zero
field and in external field.

In this Letter we present $^{69,71}$Ga NMR lineshape and
relaxation rate measurements made on Pr$_3$Ga$_5$SiO$_{14}$ as a
function of temperature in applied fields in the range $2-17$ T.
The objective is to follow the spin dynamical behavior with
temperature in the spin liquid regime and to determine how the
dynamical behavior depends on large applied magnetic fields. We
find that spin-spin relaxation rate is sensitive to field-induced
short-range magnetic ordering. This enables us to separate
frustration-induced cooperative phenomena from single-ion physics.

Single crystal samples of Pr$_3$Ga$_5$SiO$_{14}$ were grown using
the traveling floating-zone technique and characterized by X-ray
diffraction as described previously \cite{15}. The NMR spectra and
relaxation rate measurements were obtained using a pulsed
spectrometer with quadrature detection.

Figure~\ref{fig:1} shows the NMR field sweep spectrum obtained by
integrating spin-echo signals. Multiple peaks are found
corresponding to three non-equivalent sites for the $I=3/2$
$^{69}$Ga  and $^{71}$Ga isotopes each having quadrupolar
splittings which give rise to a central line and two satellites
due to non-cubic site symmetry. The resulting spectrum consisting
of eighteen overlapping lines is similar to that of
Nd$_3$Ga$_5$SiO$_{14}$ \cite{17}. Two Ga sites lie between the
kagom\'{e} planes and site 3 is randomly occupied by Ga$^{3+}$ and
Si$^{4+}$ ions \cite{17}. The principal components of the spectrum
are denoted as C(1) and C(2) for each isotope.

\begin{figure}[tbp]
\linespread{1}
\par
\includegraphics[width=3.2in]{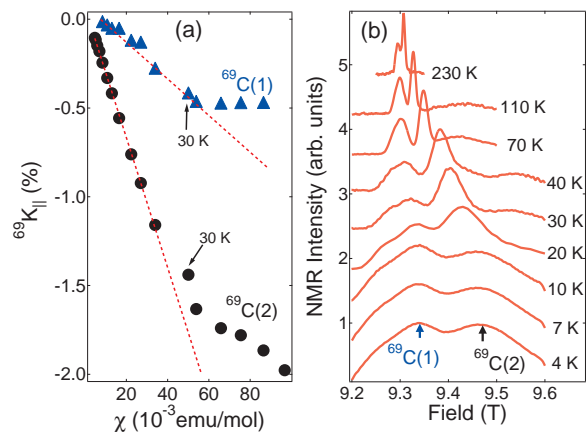}
\par
\caption{(color online). (a) $^{69}$Ga NMR shifts measured in an
applied field of 9 T along the crystal \emph{c}-axis plotted
versus the magnetic susceptibility with temperature as the
implicit parameter. For $T < 30$ K the plot shows departures from
a simple linear relationship. (b) $^{69}$Ga field-swept NMR
spectra at 95 MHz as a function of \emph{T}.} \label{fig:2}
\end{figure}

Figure~\ref{fig:2}(a) plots the measured spectral NMR shift
$^{69}K$ for $\emph{H} = 9$ T along \emph{c} versus
$\chi_{\parallel}$ with \emph{T} as the implicit parameter while
Fig.~\ref{fig:2}(b) shows the spectra in a stacked plot. The
linewidth increases significantly as \emph{T} is lowered below 100
K reaching a plateau value of 0.22 T below 10 K. For $T > 30$ K a
linear relationship between $^{69}K$ and $\chi_{\parallel}$ is
found. However, departures from this relationship become apparent
below 30 K where spin correlation effects become increasingly
important.

\begin{figure}[tbp]
\begin{center}
\linespread{1}
\par
\includegraphics[width=3.1in]{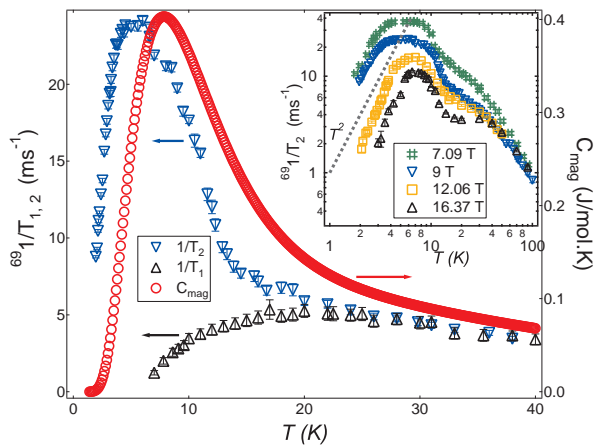}
\par
\caption{(color online). $^{69}$Ga $1/T_{1}$ and $1/T_{2}$ for
Pr$_3$Ga$_5$SiO$_{14}$ as a function of \emph{T} at 9 T, together
with magnetic specific heat at 9 T whose peak is close to the
maximum of $^{69}1/T_{2}$. The similarity in behavior of
$^{69}1/T_{2}$ and $C_{mag}$ is striking and points to a common
underlying mechanism. Inset: log-log plot of $^{69}1/T_{2}$ at
different fields. Notice that the broad maximum sharpens as the
field is increased and below the peak the behavior is close to
$T^{2}$ shown as the dotted line.} \label{fig:3}
\end{center}
\end{figure}

\begin{figure}[tbp]
\begin{center}
\linespread{1}
\par
\includegraphics[width=3.1in]{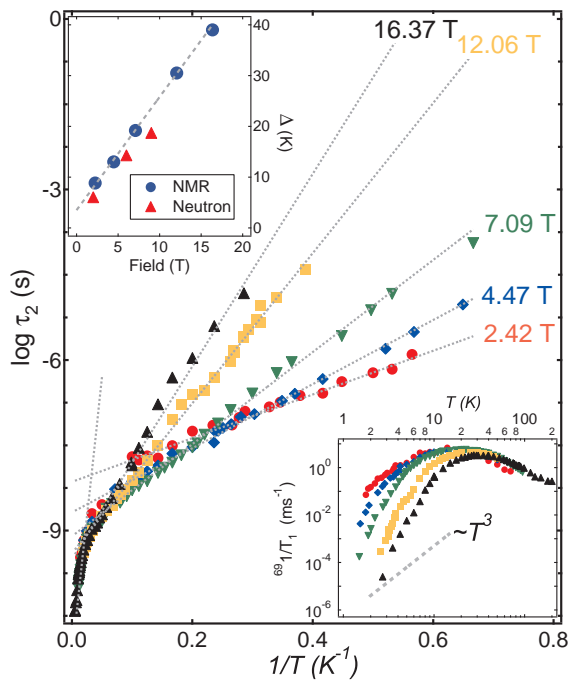}
\par
\caption{(color online). Temperature dependence of the transverse
spin correlation time $\tau_{2}$ at different fields extracted
from $^{69}1/T_{1}$ vs $T$ plot (see lower inset). At high
temperatures, the gap $\Delta \approx 98$ K obtained from the
slope is field-independent. Below 30 K, the gap
$\mathrm{\Delta_{NMR}}$ has field dependence (see upper inset)
where the dashed line corresponds to the fit
$\mathrm{\Delta_{NMR}=\Delta_{0}+\alpha H}$ with
$\alpha=g\mu_{B}=3.32\mu_{B}$ and $\mathrm{\Delta_{0}=3.5}$ K. The
field-dependence of the spin gap in the excitation spectrum as
derived from 35 mK inelastic neutron scattering results (Ref.
\cite{15}) is shown for comparison.}\label{fig:4}
\end{center}
\end{figure}

In order to study the Ga$^{3+}$ ion spin dynamics, spin-lattice
($1/T_1$) and spin-spin ($1/T_2$) relaxation rates were measured
as functions of temperature for the central $^{69}$Ga spectral
component corresponding to $^{69}$C(1) (see Fig.~\ref{fig:2}(b)).
Relaxation rate measurements were made in several different
applied magnetic fields directed parallel to the \emph{c}-axis and
the results are given in Figs.~\ref{fig:3} and~\ref{fig:4}.
Similar results were found for the other spectral components. At
low \emph{T} the magnetization recovery curve showed stretched
exponential form $M=M_{0}[1-\exp(-t/T_{1})^{\beta}]$ where
$\beta\approx 0.7$ indicating a distribution of relaxation rates.
In contrast to the sister compound Nd$_3$Ga$_5$SiO$_{14}$
\cite{17} and other transition metal based frustrated 2D
antiferromagnets such as NiGa$_2$S$_4$~\cite{11}, in which
wipe-out of the NMR signal occurs at low temperatures due to low
frequency magnetic fluctuations, the NMR signal in
Pr$_3$Ga$_5$SiO$_{14}$ could be observed in all applied fields
over the temperature range 300~mK to 290~K. As a result, the
relaxation rates could be measured over the entire temperature
range showing well defined maxima below 30 K, as seen in
Figs.~\ref{fig:3} and~\ref{fig:4}, permitting detailed analysis of
the data.

As noted above, magnetic contributions to the specific heat are
accounted for by assuming low-lying crystal field split states for
the Pr$^{3+}$ ions \cite{15}. Following this model, we may expect
that, for $T<30$ K, the temperature dependence of the correlation
time due to transitions between the ground state and the first
excited state should be given by $\tau = \tau_{0}e^{\Delta_{1}/T}$
with $\Delta_{1}$ the lowest energy gap and the pre-exponential
factor $\tau_{0} \sim 10^{-11}$ s. This predicts that for $T <
\Delta_{1}$ the correlation time for spin fluctuations will
increase rapidly with decreasing \emph{T}. Behavior of this kind
has been found in $\mu$SR and Ga NQR relaxation rate measurements
in the quasi-2D AF NiGa$_2$S$_4$ \cite{11}. However, the present
relaxation rate behavior cannot be accounted for using the
Arrhenius expression with a field-independent energy gap. The
slope of the low \emph{T} region of the log-log plot of $1/T_{1}$
versus \emph{T} in Fig.~\ref{fig:3}~(inset) gives a
field-dependent gap $\Delta_{NMR}$ which is discussed below.

Spin-lattice relaxation is attributed to fluctuating hyperfine
fields, produced by the electron moments on nearby Pr$^{3+}$ ions,
which induce nuclear spin state transitions. The fluctuating local
field $H_{L}$ at a Ga nuclear site is mainly due to dipolar
interactions with neighboring electron spins and any transferred
hyperfine interaction plays a subsidiary role. The dipolar
interaction induces nuclear transitions at the Larmor frequency
$\omega_{I}$ while an isotropic hyperfine interaction will induce
mutual electron-nucleus transitions at the much higher frequency
$|\omega_{I}\pm\omega_{S}|$ where $\omega_{S}$ is the electron
frequency. For long correlation times the dipolar process is
dominant and we can neglect any contributions due to the hyperfine
coupling.  Assuming that the spin correlation function decays
exponentially, the spin-lattice relaxation rate may be written as
$1/T_{1}=C_{\perp}(\frac{\tau_{2}}{1+\omega_{1}^{2}\tau_{2}^{2}})$
where $\tau_{2}$ is the transverse correlation time for electron
spins and $C_{\perp}=\gamma_{I}^{2}\langle H_{\perp}^{2}\rangle$
is the transverse component of the local field \cite{18,19}.
Similarly, we have
$1/T_{2}=\frac{1}{2}C_{\perp}(\frac{\tau_{2}}{1+\omega_{1}^{2}\tau_{2}^{2}})+
C_{\parallel}\tau_{1}$ with $\tau_{1}$ being the longitudinal
correlation time, $C_{\parallel}=\gamma_{I}^{2}\langle
H_{\parallel}^{2}\rangle$, and $H_{\parallel}$ the z-component of
the local field. The expression for $1/T_{2}$ holds provided
$\tau_{1}<1/\Delta\omega$ where $\Delta\omega$ is the NMR
linewidth of the selected spectral component. For $\tau_{1}\leq
1/\Delta\omega$, $1/T_{2}$ is expected to saturate. The
introduction of both transverse and longitudinal electron spin
correlation times allows for the possibility of different
mechanisms being of dominant importance for $\tau_{1}$ and
$\tau_{2}$, respectively. A maximum in $1/T_{1}$ occurs for
$\omega_{I}\tau_{2}=1$ while for higher \emph{T}, in the short
correlation time case ($\omega_{1}\tau_{2} < 1$) the dipolar
mechanism leads to $1/T_{1} \sim C_{\perp}\tau_{2}$ and $1/T_{2}
\sim \frac{1}{2}C_{\perp}\tau_{2}+C_{\parallel}\tau_{1}$.
Inspection of Fig.~\ref{fig:3} shows that for $T
> 30$ K, the $1/T_{1}$ and $1/T_{2}$ curves lie close together suggesting that the
transverse fluctuations are of dominant importance, corresponding
to $C_{\perp} > C_{\parallel}$, for both relaxation rates in this
interval. It is likely that $\tau_{1}=\tau_{2}$ at high \emph{T}.
Use of the condition for the maximum in $1/T_{1}$ permits
$\tau_{2}$ values to be obtained as a function of \emph{T} as
shown in Fig.~\ref{fig:4}. Assuming the Arrhenius relation holds
in the high-\emph{T} ($T>30$ K) region the fitted curve gives
$\tau_{0}\sim 10^{-11}$ s and the energy gap for spin excitations
$\Delta = 98$ K. This is consistent with the crystal field
splittings ($\Delta_{1}+\Delta_{2}$) given by the specific heat
results \cite{15}. The energy gap obtained from the slopes of the
curves in the low \emph{T} region of Fig.~\ref{fig:4}, denoted
$\Delta_{NMR}$, is clearly field-dependent with values plotted
versus \emph{H} in the upper inset. The fitted curve in this plot
has the form $\Delta_{NMR} = \Delta_{0} + \alpha H$ where the
slope $\alpha \approx g\mu_{B}$ with $\mu_{B}$ the Bohr magneton
and $g = 3.32$ close to the $g$ value for the Pr$^{3+}$ ion. The
magnitude of the zero-field gap, $\Delta_{0}=3.5$ K (see
Fig.~\ref{fig:4} caption), obtained from the low temperature ($T <
10$ K) NMR relaxation data is much smaller than the high-\emph{T}
NMR value or that from the specific heat results. We note that
integrated inelastic neutron scattering data at 35 mK \cite{15}
give a field - dependent spin gap in the excitation spectrum
strikingly similar to the gap derived from the NMR correlation
time behavior as shown in the upper inset in Fig.~\ref{fig:4}. It
is likely that field-suppressed magnetic fluctuations are
responsible for the observed field-dependence.

While $1/T_{1}$ decreases at temperatures below the maximum shown
in Fig.~\ref{fig:3}, $1/T_{2}$ continues to increase as \emph{T}
is lowered before passing through a maximum and then decreasing
dramatically below 7 K. The behavior is strongly field-dependent
and this again points to field suppression of magnetic
fluctuations. Fig.~\ref{fig:3} compares the behavior of $1/T_{2}$
at 9 T with that of the specific heat at 9 T with the scales
adjusted to allow comparison. The magnetic specific heat
$C_{mag}(T)$, which is obtained from the measured $C_{P}(T)$ by
subtracting the lattice contribution using
La$_{3}$Ga$_{5}$SiO$_{14}$ as a reference, shows $T^{2}$ behavior
for $H = 0.15$ T. The entropy saturates at $R\ln{3}$ as expected
for a system with three low-lying crystal field states. The
temperature dependence of $1/T_{2}$ is strikingly similar to that
of $C_{mag}(T)$ as seen in Fig.~\ref{fig:3} (for H = 9 T) and this
similarity in form points to a common underlying process involved
in both observed behaviors linked to the spins entering the
singlet ground state at low \emph{T}. Similar behavior is observed
in NiGa$_2$S$_4$ where the estimated spin ``freezing'' temperature
T$_{f}$ is coincident with the peak in C$_{mag}$, but NMR wipeout
prevents detailed comparison with $1/T_{2}$ \cite{11}.
Pr$_3$Ga$_5$SiO$_{14}$, however, allows this comparison to be made
over a complete temperature range. Below 7 K occupation of the
singlet ground state by the electron spins rapidly increases with
decreasing \emph{T} and both $1/T_{2}$ and $C_{mag}$ drop to low
values. The behavior of $1/T_{2}$ below 7 K may be accounted for
by decreases in $\tau_{1}$ and/or $H_{\parallel}$. While it is
unlikely that $\tau_{1}$ will shorten at low \emph{T}, local
ordering of spins could result in a reduction in $H_{\parallel}$
leading to a change in the ``static'' field contribution to
$1/T_{2}$. This effect will be particularly marked if spin
ordering occurs in the a-b plane. The NMR spectra for $T < 10$ K
show little change in linewidth and the Knight shift is roughly
constant (Fig.~\ref{fig:2}).

The quadratic \emph{T}-dependence of $C_{mag}$  for $H = 0$ T is
interpreted as evidence for gapless Goldstone modes. For $H > 0$
T, $C_{mag}(T)$ has a minimum at 80 mK with the low-\emph{T}
upturn ascribed to nuclear contributions to the specific heat
\cite{15}. Elastic neutron scattering measurements for $T < 1$ K
show that nanoscale ordering occurs in the presence of an applied
field consistent with 2D short-range order and give a correlation
length of 29 $\mathrm{\AA}$ ($\sim$ 6 to 7 in-plane lattice
spacings) for $H = 9$ T \cite{15}. The present relaxation rate
results suggest that at low temperatures the spin correlation time
$\tau$ becomes long as shown in Fig.~\ref{fig:4} and a possible
reduction in the \emph{c}-axis component of the dipolar field at
Ga sites results from in-plane short range ordering leading to the
anomalous behavior of $1/T_{2}$.

In conclusion, we have investigated the spin dynamics in the
rare-earth kagom\'{e} system Pr$_{3}$Ga$_{5}$SiO$_{14}$ via
$^{69}$Ga NMR measurements over a complete temperature range
without the NMR wipeout effect seen in the sister compound
Nd$_{3}$Ga$_{5}$SiO$_{14}$. The correlation time $\tau$ for spin
fluctuations extracted from the spin-lattice relaxation rate
values exhibits novel features; $\tau$ increases with decreasing
\emph{T} and below 30 K the results are consistent with a
field-dependent energy gap in the excitation spectrum. The
spin-spin relaxation rate shows a maximum close to that in the
specific heat and the form of the temperature dependence of these
two quantities is very similar below 10 K. This result warrants
theoretical attention. The drop in $1/T_{2}$ below 7 K is
attributed to a decrease in the local field at Ga sites linked to
field-induced nanoscale ordering of the electron spins in the a-b
plane.





\begin{acknowledgments}
This work was supported in part by NSF DMR-0602859 and performed
at the National High Magnetic Field Laboratory, which is supported
by NSF Cooperative Agreement No. DMR-0084173, EIEG grant, by the
State of Florida, and by the DOE.
\end{acknowledgments}

\end{document}